\documentclass{article}
\newcommand{\be}{\begin{equation}}
\newcommand{\ee}{\end{equation}}
\newcommand{\bea}{\begin{eqnarray}}
\newcommand{\eea}{\end{eqnarray}}
\newcommand{\nn}{\nonumber}
\usepackage{amsmath}
\usepackage{amsfonts}
\usepackage{amssymb}
\begin{document}
{\center {\bf Quantization of Scalar Field Theory with Internal Symmetry}\\
\vspace{10mm}

A. V. Shurgaia\footnote{e-mail: avsh@rmi.ge} \\
\vspace{5mm}
Department of Theoretical Physics \\
A. Razmadze Mathematical Institute  \\
  0193 Tbilisi \\
             Georgia \\}




\begin{abstract}
A simple theoretical model of scalar fields in one spatial dimension with internal symmetry is considered. Assuming the existence of localized classical field configurations, the Schr\"{o}dinger picture is used to describe their quantum properties. Using the collective coordinates method for the Schr\"{o}dinger equation allows the development of a perturbation theory that accurately describes the symmetry properties of the theory. As examples, the symmetries $U(1)$ and $SU (2)$ are analyzed and the discreteness of the energy of bound states is shown as a result of the symmetry of the theory.   \\

\end{abstract}
\vspace{15mm}
\section{Introduction}

Quantization of localized solutions in space  has a long history. There are several approaches to this problem, mainly within the framework of a
functional integration\cite{dash}-\cite{shur}. The method of collective coordinates is one of them. We apply in the present investigation the operator approach of this method for solving the Schr\"{o}dinger equation. The advantage of  approach is the absence of an ambiguity related with operator ordering. Besides, a perturbation theory can be constructed that is manifestly invariant with respect to the group of the symmetry of a theory under consideration in every order of the perturbation series.

We consider below the simple model with internal symmetry, which is of interest even today, although many articles were devoted to this issue\cite{tom},\cite{raj}-\cite{mon}. The point is that it is possible in theories with internal symmetry to obtain stable field configurations in more than one space dimension. In papers\cite{tom,raj,mon} the charge symmetry has been studied and contribution of the charge depending terms into the energy of the system has been evaluated. We want to focus our attention on\cite{raj}. In this article the charge symmetry is investigated in detail. It has been shown, that there exist charged classical field configurations  that are stable owing to the existence of a charge. The centrifugal term and its contribution to the  energy  have been calculated in the framework of the functional integration. The charge is assumed to take on integer values.
The extension of the method to the case of $SU(2)$ symmetry is discussed. The approach used in the present article starts from the quantum theory and suggests the construction of a perturbation theory around the classical neutral field configuration (around charged configurations is also possible). The charge (or the isotopic spin) is quantized such that the exact dependence of the energy on the charge (or the isotopic spin) is represented. Besides, the operator ordering problem does not arise in this approach.

\section{The model with the $U(1)$symmetry}

    We consider the model of scalar triplets in one space dimension with a broken $SU(2)$ symmetry. The Hamiltonian of the model is
\bea
H=\int dx\{\frac{1}{2}\pi_\alpha(x) \pi_\alpha(x) + \frac{1}{2}\frac{\partial\varphi_\alpha(x)}{\partial x} \frac{\partial\varphi_\alpha(x)}{\partial x}+ U(\varphi_i(x),\varphi_3(x),g)\}
\eea
with the following commutation relations between the field operators:
\bea
[\pi_\alpha(x),\pi_\beta(y)]=i\delta_{\alpha \beta}\delta(x-y).
\eea
the Greek indices $\alpha,\beta,\gamma...$  take on values $1,2,3$ and Latin indices - $1,2$. The fields $\varphi_i(x)$ form the charged fields. The potential $U(\varphi_i(x),\varphi_3(x),g)$ is suggested to be breaking the $SU(2)$ symmetry and obeys the condition:
\bea
U(\varphi_i(x),\varphi_3(x),g)=g^2U(g\varphi_i(x),g\varphi_3(x),1).
\eea
Thus the model is $U(1)$ invariant and as a result the charge of the system
\bea
Q=\int dx \{\varphi_1(x)\pi_2(x)-\varphi_2(x)\pi_1(x)\}
\eea
is conserved. We apply the method of collective coordinates to the Schr\"{o}dinger equation
\bea
H\Psi(\varphi_i(x),\varphi_3(x))=E\Psi(\varphi_i(x),\varphi_3(x)).
\eea
mention that the operator $\pi_\alpha(x)$ is considered as the functional derivative $\delta/{i\delta\varphi_\alpha(x)}$.

Let us introduce the transformation:
\bea
\varphi_i(x)=D_{ij}(\vartheta)(gu(x)\delta_{j1}+\Phi_j(x)),\qquad \varphi_3(x)=g\sigma(x)+\Phi_3(x),
\eea
in which the parameter $\vartheta$  together with $\Phi_\alpha(x)$ compose the new set of operators and the matrix $D(\vartheta)$ is the known two dimensional  matrix of a rotation:
\bea
D=\left(
  \begin{array}{cc}
    \cos\theta & -\sin\theta \\
    \sin \theta& \cos\theta \\
  \end{array}
\right)
\eea
In order to retain the total number of independent variables a subsidiary condition has to be imposed,
namely
\bea
\int dx N(x)\Phi_2(x)=0.
\eea
The c-number $N(x)$ is normalized as follows:
\bea
\int dx N(x)u(x)=1.
\eea
We should now express the momenta $\pi_\beta(y)$ in terms of the new variables $\theta$, $\Phi_\alpha(x)$.
This is done according to the ordinary rules of differentiation provided that the subsidiary condition (8) will be taken into account. In addition to (8) we need to introduce the following projection operator: $A_{ij}(x,y)=\delta_{ij}(x-y)-\delta_{i2}\delta_{j2}N(x)u(x)$ with the properties:
\bea
\int dy A_{i2}(x,y)N(x)=\int dx u(x)A_{2j}(x,y)=0.
\eea
The fields $\Phi_i(x)$ can be expressed as
\bea
\Phi_k(y)=\int dz A_{kj}(y,z)\overline{D}(\vartheta)_{ji}(\vartheta)(\varphi_i(z)-\delta_{i1}u(z)),
\eea
in which $\overline{D}(\vartheta)D=I$.
Taking the functional derivative of (8) and (11) with  respect to $\varphi_i(x)$, one obtains:
\bea
  &&\pi_i(x)=\overline{D}_{ji}(\vartheta)\{\Pi_j(x)+\frac{N(x)\delta_{2j}}{g-F}(p_\vartheta+\overline{Q})\} \\
  &&\pi_3(x)\equiv\Pi_3(x)=\frac{\delta}{i\delta\Phi_3(x)}
\eea
The $U(1)$ symmetry is the simplest internal symmetry.
In this expression the following notations are used:
$$\overline{Q}=\overline{J}_{ik}\int dx \Phi_k(x)\Pi_i(x),$$ and is the charge of mesons (described by fields $\Phi_i(x)$), the antisymmetric matrix $$\overline{J}_{ik}=\frac{d\overline{D}_{ik}}{d\vartheta}\mid_{\theta=0}$$ and is the element of the algebra of the group $U(1)$, $F=\int dx N(x)\Phi_1(x)$.  For the new set of the variables of the system the following nonzero commutation relations hold:
\bea
[\vartheta,p_\vartheta]=i,\;[\Phi_i(x),\Pi_k(y)]=iA_{ik}(x,y),\; [\Phi_3(x),\Pi_3(y)]=i\delta(x-y).
\eea
    We now can write in the new representation the kinetic energy:
\bea
    &&K=\frac{1}{2}\int dx\pi_a(x)\pi_a(x)= \nn \\
&&\frac{1}{2}\int dx\{\Pi_3^2(x)+[\Pi_k(x)+\frac{N(x)\delta_{k2}}{g-F}(p_\vartheta+\overline{Q})] \nn
  [\Pi_k(x)+ \\ &&\frac{N(x)\delta_{k2}}{g-F}(p_\vartheta+\overline{Q})] -i\frac{N(x)\Pi_1(x)}{g-F}\}
\eea
As we can see the kinetic energy does not contain the  variable $\vartheta$. It is easy to verify that in the new representation  the charge operator $Q$ is reduced to $p_\vartheta:\;Q=p_\vartheta$. The equality (3) and the transformation (6) make it possible to expand the potential $U$ into series in inverse powers of $g$, namely
\bea
U=g^2U(u(x),\sigma (x))+g\frac{\partial U(u(x)),\sigma (x)}{\partial u(x)}+g\frac{\partial U(u(x)),\sigma (x)}{\partial \sigma(x)}+ \nn \\
\frac{1}{2}\frac{\partial^2U(u(x)),\sigma (x)}{\partial u(x)^2}+\frac{1}{2}\frac{\partial^2U(u(x)),\sigma (x)}{\partial \sigma(x)^2}+\frac{\partial^2U(u(x),\sigma (x)}{\partial \sigma(x)\partial u(x))}+...
\eea
Thus the variable $\vartheta$ is cyclic and the operator $p_\vartheta$ can be replaced with the c-number. Before doing this, we introduce the transformations , which eliminate the linear and cross terms in the operators $\Pi_i$ from the kinetic energy. Firstly we assume $N(x)=\lambda u(x)$ such that owing to (9) $$\lambda \int dx u^2(x)=1.$$ The quantity $F$ can now be rewritten as $\lambda h$ with $h=\int dx u(x)\Phi_1(x)$. Secondly we change the wave functional $\Psi(\vartheta , \Phi_\alpha (x))$ as follows:
$$
\Psi(\vartheta , \Phi_\alpha (x))=\frac{1}{\sqrt{g-\lambda h}}\widetilde{\Psi}(\vartheta , \Phi_\alpha (x)).
$$
Taking functional derivative with respect to $\Phi_1(x)$ one obtains for the Hamiltonian of the system
\bea
&&H=\int dx \{\frac{1}{2}\frac{\partial}{\partial x}(gu(x)\delta_{i1}+\Phi_i(x))
\frac{\partial}{\partial x}(gu(x)\delta_{i1}+\Phi_i(x))+ \nn \\
&&\frac{\partial}{\partial x}(g\sigma (x)+\Phi_3(x))\frac{\partial}{\partial x}(g\sigma (x)+\Phi_3(x))+U+\frac{1}{2}\Pi_\alpha(x)\Pi_\alpha(x)\}+ \nn \\
&&\frac{\lambda}{2}[\frac{1}{g-\lambda h}(p_\vartheta+\overline{Q})\frac{1}{g-\lambda h} (p_\vartheta+\overline{Q})-\frac{1}{4}]
\eea
which acts now on $\widetilde{\Psi}(\vartheta,\Phi_\alpha (x))$. The operator $U$ is assumed to be expanded as in (16). We can now factor out the $\vartheta$-dependence in the wave functional since $\vartheta$ is cyclic:
$$
\widetilde{\Psi}(\vartheta,\Phi_\alpha (x))=exp(im\vartheta)\Psi'(\Phi_\alpha(x))
$$
with $m=0,\pm 1,\pm 2,\pm 3...$ replacing the operator $p_\vartheta$ with $m$. All the calculations are so far accurate and we did not make any approximation. We have eliminated the $\vartheta$-dependence from the Hamiltonian and it is now the function of the quantum number $m$, thereby the charge has been quantized. One can now construct a perturbation theory (which would be manifestly $U(1)$-invariant in every order)  expanding the Hamiltonian, the energy and the $\Psi'(\Phi_a(x)$ into series in inverse powers of $g$ as follows:
\bea
&&H=g^2H_0+gH_1+H_2+g^{-1}H_3+g^{-2}H_4+... \nn \\
&&E=g^2E_0+gE_1+E_2+g^{-1}E_3+g^{-2}E_4+...  \\
&&\Psi'(\Phi_\alpha(x))=\Psi_0+g^{-1}\Psi_1+g^{-2}\Psi_2+...\nn
\eea
We next solve the system of equations:
\bea
 &&(H_0-E_0)\Psi_0=0 ,\nn \\
 &&(H_0-E_0)\Psi_1+(H_1-E_1)\Psi_0=0,\\
 &&(H_0-E_0)\Psi_2+(H_1-E_1)\Psi_0=0+(H_2-E_2)\Psi_0=0,\nn \\
 &&..........................................................................\nn
\eea
We do not specify the potential $U$ since we aimed to quantize the charge and to obtain the exact dependence of the Hamiltonian on the charge. We only make some remarks regarding the equations (19). It is evident, that the leading equations of order $g^2$ and $g$ reproduce the classical equations of motion giving zero charge solutions for $u(x)$ and $\sigma(x)$ and classical energy of neutral field configurations. The equation of order $g^{0}$ is bilinear in field operators $\Phi_\alpha(x)$ and $\Pi_\alpha(x)$ that formally can be diagonalized. The result is infinite sum of oscillators, which must be regularized. The charge dependence of the energy arises in the approximation of the order $g^{-2}$ and is of the form: $$E_4=\frac{\lambda}{2}(m^2+<\overline{Q}^2>-\frac{1}{4}).$$ The symbol $<>$ denotes  the average over the ground  states of oscillators.

\section {$SU(2)$ symmetry}

In this section we apply the method to the same model, but with $SU(2)$ symmetry. The fields $\varphi_\alpha(x)$ are now assumed to belong to the adjoin representation of  $SU(2)$.  The corresponding conserved quantity is an isotopic spin, the third component of which defines the charge of a system:
\bea
t_\alpha =\varepsilon_{\alpha \beta \gamma}\int dx \varphi_\beta (x)\pi_\gamma (x).
\eea
The transformation which introduces collective coordinates is similar to (6) but with $D(\theta,\phi)$ depending on two parameters:
\bea
\varphi_\alpha(x)=D_{\alpha\beta}(\theta,\phi)\{u(x)\delta_{\beta 3}+\Phi_\beta (x)\}.
\eea
So the phase space is now extended by two additional variables and together with the fields $\Phi_\alpha(x)$ they form the new set of variables. The subsidiary conditions and the projection operator needed for evaluating the momentums in the new representation are:
\bea
&&\int dx N_{ik}\Phi_k(x)=0, \\ &&A_{ij}=\delta_{ij}\delta(x-y)-N_{ik}(x)M_{kj}(y).
\eea
The quantity $N_{ik}$ can be chosen to satisfy the condition
\bea
\int dx N_{ik}(x)M_{kj}(x)=1.
\eea
Omitting the details of calculation, we give the final expression for momenta $\pi_\alpha (x)$:
\bea
\pi_\alpha (x)=\overline{D}_{\beta\alpha}(\theta,\phi)\{\Pi_{\beta}(x)-(g+F)^{-1}_{j\sigma}N(x)_{ji}\delta_{\beta i}[\overline{t}_{\sigma}+T_{\sigma}]\}
\eea
in which the following notations are used\footnote{Greek indices take on values 1,2,3 and Latin indices - 1,2.} :
\bea
&&F_{\sigma j}=\int dx (\overline{J}_{\sigma})_{i\nu}N_{ji}(x)\Phi_{\nu}(x), \\
&&\delta_{\sigma m}M_{im}(x)=\delta_{\sigma m}(\overline{J}_{m})_{i3}u(x),  \\
&&\overline{T}_{\sigma}=(\overline{J}_{\sigma})_{i\nu}\int dx \Phi_{\nu} (x)\Pi_{i}(x),
\eea
The matrices $(\overline{J}_{\sigma})_{\mu\nu}=\varepsilon_{\sigma\mu\nu}$ are the elements of the algebra of the adjoint representation of $SU(2)$. The matrix
\bea
{D}(\theta,\phi)=\left(
  \begin{array}{ccc}
    \cos\theta\cos\phi & -\sin\phi &  \sin\theta\cos\phi \\
    \cos\theta\sin\phi & \cos\phi & \sin\theta\sin\phi \\
    -\sin\theta & 0 & \cos\theta \\
  \end{array}
\right)
\eea
with $\overline{D}d=1$. The nonzero commutation relations of new variables are:
\bea
[\theta,p_\theta]=[\phi,p_\phi]=1,\;[\Phi_i(x),\Pi_j(y)]=A_{ij}(x-y).
\eea
The operator $\overline{t}_\alpha$ are the generators of the inverse transformations of $SU(2)$, whereas $t_\alpha$ the generators of  direct transformation and $$t_\alpha=D_{\alpha2}p_\theta-D_{\alpha1}\frac{1}{\sin\theta}p_\phi.$$ These two operators are related by equality $t_\alpha=\overline{D}_{\alpha\beta}t_b$. he further simplification can be introduced by setting $N_{ji}=\lambda(\overline{J}_{m})_{i3}u(x)$ such that the equality is replaced by
\bea
\lambda\int dx u^2(x)=1.
\eea
Besides, $F_{\sigma j}=-\lambda\int dxu(x)\Phi_3(x)=-\lambda h$.
As in the case of $U(1)$ symmetry we assume the   wave functional to be $$\Psi=(g-\lambda h)^{-1}\widetilde{\Psi}.$$
After all this one obtains the following expression for the total Hamiltonian of the model under consideration:
\bea
&&H=\int dx \frac{1}{2}\{\Pi_\alpha(x)\Pi_\alpha(x)+ (gu'(x)\delta_{\alpha3}+
\Phi_\alpha'(x)) (gu'(x)\delta_{\alpha3}+\Phi_\alpha'(x))+U\}\nn \\
&&\frac{\lambda}{2}(g-\lambda h)^{-1}\{\overline{t}_{j}+\overline{T}_{j})(g-\lambda h)^{-1}\{\overline{t}_{j}+\overline{T}_{j}).
\eea
The potential $U$ is represented as series similar to the case of $U(1)$ symmetry  and we do not give corresponding relation.
One can easily verify the correctness of the equality $T_\alpha=t_\alpha$ and since the commutation  relation $[t_\alpha,\overline{t}_\beta]=0$ holds the operators $\overline{t}_{j}$ can be replaced by c-numbers. The perturbation theory is similar to that of in preceding section. It is evident that the dependence of the energy from the quantum number of isotopic spin is of order $g^{-2}$  and the energy in this order is proportional to $j(j+1)$ with integer $j.$

\end{document}